\documentstyle[12pt]{article}

\large
\newcommand{\be}{\begin{eqnarray}}
\newcommand{\ee}{\end{eqnarray}}

\begin{document}
\setlength{\baselineskip}{21pt}
\pagestyle{empty}
\vfill
\eject
\begin{flushright}
SUNY-NTG-94/6
\end{flushright}

\vskip 2.0cm
\centerline{\bf Spectrum of the Dirac operator in a QCD instanton liquid:}
\centerline{\bf two versus three colors}
\vskip 2.0 cm
\centerline{Jacobus Verbaarschot}
\vskip .2cm
\centerline{Department of Physics}
\centerline{SUNY, Stony Brook, New York 11794}
\vskip 2cm

\centerline{\bf Abstract}
Approximating the sum over all gauge field configurations in the QCD
partition function by a liquid of instantons, we calculate the spectrum
of the Dirac operator for two and three colors and for 0, 1 and 2 flavors.
We find a remarkable difference in the spectrum near zero virtuality between
2 and 3 colors, which can be  explained in terms of chiral random
matrix theory. For two colors the Dirac operator is real, and the appropriate
random matrix ensemble has real matrix elements. For three colors the Dirac
operator is complex, and the spectrum can be  described by a
random matrix ensemble with complex matrix elements.
These results provide further evidence  that the spectrum
near zero virtuality is universal and is completely
determined by symmetries.

\vfill
\noindent
\begin{flushleft}
SUNY-NTG-94/6\\
February 1994
\end{flushleft}
\eject
\pagestyle{plain}

\vskip 1.5cm
\noindent{\bf 1. Introduction}
\setcounter{equation}{0}
\vskip 0.5 cm
It is widely believed that at some critical temperature
the phase of  QCD changes from a chirally broken phase to a chirally
symmetric phase.
The order parameter is the chiral condensate, which, according
to the Banks-Casher formula \cite{BANKS-CASHER-1980}, is proportional to the
thermodynamic limit of the spectral density of the
Dirac operator at zero virtuality.
The study of this order parameter is the central objective
of this paper. In particular, we will analyze the spectrum
$near$ zero virtuality
via the microscopic limit which is obtained in the thermodynamic limit
$V_4\rightarrow \infty$ by rescaling the eigenvalues by $V_4$.

A natural way to perform these studies would be via lattice QCD calculations.
However, the order parameter is defined in the chiral limit which makes such
calculations with dynamical fermions infeasible. Instead, we
approximate the integral over all gauge field configurations
by a liquid of
instantons and anti-instantons. In this approach
the chiral limit does not give rise to technical problems.

Our main goal
is to substantiate the claim \cite{SHURYAK-VERBAARSCHOT-1993}
that spectral correlations on the scale of
one average level spacing are universal. It is well known from the study
of systems that are classically chaotic
\cite{BOHIGAS-GIANNONI-1984,SELIGMAN-VERBAARSCHOT-ZIRNBAUER-1984}
\cite{SELIGMAN-VERBAARSCHOT-1985,MEHTA-1991}
\cite{SIMONS-SZAFER-ALTSCHULER-1993}
that correlations between
eigenvalues measured in units of the average level spacing are universal
in the sense that they are solely determined by the symmetries of the
system. Therefore they can as well be described by random matrix theory
with only the symmetries as input. It is our conjecture
that the same is true for correlations of the eigenvalues of
the Dirac operator measured over the ensemble of all gauge field
configurations. In QCD, though, we have an important additional symmetry:
the chiral symmetry. As a consequence all nonzero eigenvalues occur in pairs
$\pm \lambda$. Obviously, this symmetry has to be implemented in the
random matrix model. What is important is that for these types of
random matrix models, which we call chiral random matrix models, the
point $\lambda = 0$ is special: the spectrum has a hard edge at
$\lambda = 0$ \cite{FORRESTER-1993,TRACY-WIDOM-1993}.
Therefore, correlations on the scale of one average level spacing
can be defined with respect
to this point. The microscopic spectral density is exactly such
correlator and is expected to be universal. There are other arguments
in favor of universality. Most notably, it was recently argued by Leutwyler
and Smilga \cite{LEUTWYLER-SMILGA-1993}
that the low-energy effective partition function of QCD which is
solely based on symmetries leads to an infinite family of sum rules
for the inverse powers of the eigenvalues of the Dirac operator. Since
all sum rules can be derived from the microscopic correlators, this strongly
suggest that the spectrum at this scale is determined by symmetries
as well.

For the classical random matrix ensembles we have three universality classes:
the GOE, the GUE and the GSE, depending on whether the matrix elements
are real, complex or quaternion real, respectively \cite{DYSON-1962}.
The same is true
for the chiral random matrix ensembles. Recently, we succeeded the classify
the $SU(N_c)$ gauge theories accordingly \cite{VERBAARSCHOT-1994P}.
For $N_c =2$ and fundamental
fermions all matrix elements of the Dirac operator can be made real,
for $N_c \ge 3$ gauge groups with fundamental fermions the Dirac operator
is complex, whereas for adjoint fermions, the Dirac operator can be
regrouped into quaternions. The corresponding random matrix ensembles
will be called the chiral orthogonal ensemble (chGOE), the chiral unitary
ensemble (chGUE) and the chiral symplectic ensemble (chGSE), respectively.
Each of the three cases can be identified with one
of the three possible scenarios of chiral symmetry
breaking \cite{DIAKONOV-PETROV-1993}.
Up to now, the microscopic spectral density has been obtained for the chGOE
\cite{VERBAARSCHOT-1994} and the chGUE \cite{VERBAARSCHOT-ZAHED-1993}.

As a final note in this introduction, we want to point out that
the microscopic fluctuations become universal only after they are separated
from all other variations in the spectrum. This procedure, called unfolding
\cite{BOHIGAS-GIANNONI-1984}, will be applied to all our  numerical results.
It can only be achieved if a separation of scales takes place. We want to
stress that this implicit assumption is a nontrivial point, that cannot be
taken for granted.

In this paper we calculate the spectrum of the Dirac operator for
$SU(2)$ and $SU(3)$ color gauge groups
and compare the results with the universal functions
obtained from random matrix theory. The microscopic spectral density is defined
in section 2. The main ingredient of the instanton liquid model are discussed
in section 3. The expressions for the microscopic spectral density are given in
section 4. Numerical results are presented in section 5 and concluding remarks
are made in section 6.

\vskip 1.5cm
\noindent{\bf 2. The microscopic spectral density}
\vskip 0.5 cm
For $N_f$
flavors with masses $m_f$ ($m_f \rightarrow 0$) the QCD partition function
in the sector with $\nu$ zero modes is defined by
\be
Z_\nu^{\rm QCD}(m) = \langle \prod^{N_f}_{f=1}\prod_{\lambda_n>0}
(\lambda_n^2 + m_f^2) m_f^{\nu}\rangle_{S_\nu(A)},
\ee
where the average is over gauge field configurations
with $\nu$ fermionic zero modes weighted by the gauge field action
$S_\nu(A)$. The product is over all eigenvalues of the
Dirac operator.
The spectral density of the QCD Dirac operator for a
given field configuration is defined by
\be
\rho(\lambda) = \sum_n \delta(\lambda-\lambda_n).
\ee
The average spectral density $\langle \rho(\lambda)\rangle$
is defined with respect to the partition function (1). According
to the Banks-Casher formula \cite{BANKS-CASHER-1980}
the chiral order parameter can be expressed as
\be
\langle \bar\psi\psi\rangle = \lim_{V_4\rightarrow \infty}\frac 1{V_4} \rho(0),
\ee
where it is understood that the limit $V_4 \rightarrow \infty$ is taken
before the chiral limit $m \rightarrow 0$. Its existence implies that
the eigenvalues near zero scale as $1/V_4$. Therefore,
the order parameter can be analyzed in much greater detail
via the microscopic limit of the spectral density defined as
\be
\rho_S(z) = \lim_{ V_4\rightarrow \infty} \frac 1V_4 \left
<\rho(\frac zV_4)\right >_\nu.
\ee
Instead of a single number, our new order parameter is a function that
not only gives the thermodynamic limit
 of $\langle \bar\psi\psi\rangle$, but also
provides information on $how$ the thermodynamic limit is approached.

\vskip 1.5cm
\noindent{\bf 3. Instanton liquid model}
\vskip 0.5 cm
The sum over all gauge field configurations in (1) has only been performed
exactly in lattice gauge theory. However, in calculations with dynamical
fermions, the quark masses cannot be taken equal to zero, which makes it
difficult to study the chiral limit of the microscopic spectral density.
This problem is not present in our approach, where we approximate the
sum over all gauge field configuration in (1) by a
sum over instantons and anti-instantons.

In this approximation, the non-zero mode quantum fluctuations about
the classical configurations are taken into account to one-loop order under the
assumption that contributions from different instantons factorize. This
is justified when the ensemble of instantons is sufficiently dilute,
which according to phenomenological arguments \cite{SHURYAK-1993},
is indeed the case.
The integral over the zero modes,
$i.e.$ the size, the position and the orientation of each of the
pseudoparticles, is done
exactly via a Monte Carlo simulation.
Also the integral over the fermionic zero modes is performed
exactly. The latter amounts to evaluating the fermion determinant
in the space of fermionic zero modes under the assumption that it factorizes
from the contribution of the non-zero modes.
Again, this can be justified on the basis
that the instanton ensemble is sufficiently dilute.
The resulting Dirac operator is given by
the $(N_I+N_A)\times( N_I+N_A)$ matrix
\be
\left( \begin{array}{cc}i m_f & T\\
                 T^\dagger & im_f
\end{array}\right ),
\ee
where the overlap matrix elements are given by
\be
T_{IA}
=\int d^4x \psi_0^{\dagger I}(x) i\hat D  \psi_0^A(x)=
\frac 1{2(\rho_I \rho_A)^{\frac 12}} {\rm Tr}
\left ( \tau_\mu^+ \hat R^{IA}_\mu U_I^{-1} U_A \right )  F(\lambda),
\ee
Here, $R_I, R_A$ are the positions of the instanton and the anti-instanton,
$\rho_I$, $\rho_A$, their sizes and
$U_I$ and $U_A$ their orientations in color space. The vector $\hat R^{IA}_\mu$
is a unit vector in the direction $(R_I - R_A)_\mu$, and we have introduced the
notation $\tau^+_\mu = (\vec \tau, i)_\mu$ with  $\vec\tau$ are the Pauli spin
matrices.
The scalar function
$F(\lambda)$ depends on the specific ansatz for the gauge field configuration.
We use the so called streamline field configurations
\cite{YUNG-1988,VERBAARSCHOT-1991}, which are optimal in
the sense that the first derivative with respect to directions perpendicular
to the collective coordinates vanishes. In this case $\lambda$ is given by a
conformal invariant combination of $R_{IA}$, $\rho_I$ and $\rho_A$.
Asymptotically, we have
\be
F(\lambda) \sim 4 \frac{(\rho_I\rho_A)^{3/2} }{R_{IA}^3}
\ee
At this moment we want to point out that
for $SU(2)$ all matrix elements of  $T$ are $real$, whereas for $SU(3)$
and all larger gauge groups they are $complex$. As we will see below, this
distinction has important consequences for the spectral correlations
near zero virtuality.

Summarizing, the QCD partition function is approximated by
\be
Z = \int \prod_{i=1}^N d U_i dz_i d\rho_i \mu(\rho_i\Lambda)
\prod_f^{N_f}\det(T^\dagger T + m_f^2)
\exp(-\beta(\rho\Lambda) \sum_{i < j} S^{int}_{ij}),
\ee
where $\mu(\rho\Lambda)$ is size distribution as induced by the quantum
fluctuations ($\Lambda$ is the QCD scale parameter).
The total density of instantons $N/V_4$ is kept fixed
($N = N_I + N_A$), and we always work
in the sector of zero total topological charge. For the interaction of
the instantons we use the streamline action supplemented by a hard core.
The latter has been introduced in order to be able to simulate a stable
liquid of instantons. In this way, instantons are simulated as they
occur in the full QCD partition function as given by lattice
QCD simulations \cite{HANDS-TEPER-1990}.
For a more detailed discussion of the instanton liquid model described above
we refer to refs.
\cite{DIAKONOV-PETROV-1984,SHURYAK-1988,SHURYAK-VERBAARSCHOT-1990}.

\vskip 1.5cm
\noindent
{\bf 4. Random matrix theory}
\vskip 0.5cm
Our hypothesis is that the fluctuations of the eigenvalues of the Dirac
operator near zero virtuality are universal and can be described by chiral
random matrix theory. Using the instanton partition function as an inspiration
\cite{NOWAK-VERBAARSCHOT-ZAHED-1989,SIMINOV-1991,SHURYAK-VERBAARSCHOT-1993}
it is clear that the appropriate random matrix model in the sector with $\nu$
fermionic zero modes is defined by \cite{VERBAARSCHOT-1994P}
\be
Z_{\beta,\nu}(m) = \int {\cal D}T P_\beta(T)\prod_f^{N_f}\det \left (
\begin{array}{cc} m_f & iT\\
                 iT^\dagger & m_f
\end{array}\right ),
\ee
where $T$ has the symmetries of the corresponding Dirac operator and the
masses are in the chiral limit ($m_f\rightarrow 0$). As discussed above,
depending on the universality class the matrix $T$
is real ($\beta = 1$, chGOE), complex ($\beta = 2$, chGUE)
or quaternion real
($\beta = 4$, chGSE). In the latter case
the square root of the fermion determinant appears in (9).
The matrix $T$ is a rectangular $n\times m$ matrix with $|n-m| = \nu$
(for definiteness we take $m > n$), so that the 'Dirac operator' in (9) has
exactly $\nu$ zero modes.
The function $P(T)$ is chosen gaussian:
\be
P_\beta(T) = \exp\left (
{-\frac{\Sigma^2\beta n}{2}\sum_{k=1}^n \lambda_k^2}\right).
\ee
In this normalization, the chiral condensate (as given by (3))
equals $\Sigma$ in each of the
three random matrix ensembles.
The total number of modes is $N\equiv m+n$. The latter quantity
is identified with the volume of space time.

The spectral density can be  obtained by transforming to new integration
variables in which $T$ is diagonal. For $\beta = 2$ the corresponding Jacobian
can be rewritten in terms of generalized Laguerre polynomials. After some
manipulations, which are well-known from random matrix theory
\cite{MEHTA-1991},
one obtains a formula for
the spectral density in terms of these polynomials. From
their asymptotic properties it follows that the microscopic limit is
given by
\be
\rho_S(z)  = \frac {\Sigma^2 z}{2} (J^2_{N_f+\nu}(\Sigma z) -
J_{N_f+\nu+1}(\Sigma z) J_{N_f+\nu-1}(\Sigma z)).
\ee
In a more modern language, this limit is a double scaling limit, which
exists as a consequence of the fact that the Laguerre ensemble has a
hard edge at zero \cite{FORRESTER-1993,TRACY-WIDOM-1993}.

The situation for the chGOE is much more complicated. Because of the structure
of the Jacobian, the standard orthogonal polynomial method no longer works.
However, in this case it is possible
to express the spectral density in terms of skew-orthogonal polynomials
\cite{MAHOUX-MEHTA-1991}. Remarkably, using a number of tricks from
\cite{NAGAO-WADATI-1991} it is also possible in this case to obtain
an analytical result for the microscopic spectral density
\cite{VERBAARSCHOT-1994}. The result can be expressed as
an integral over the Bessel kernel
\be
\rho_S(z) = \frac {\Sigma}{4} J_{2a+1}(z{\Sigma}) &+& \frac {\Sigma}{2}
\int_0^\infty dw (zw)^{2a+1} \epsilon(z-w)
\left ( \frac 1w \frac d{dw} - \frac 1z \frac d{dz}\right )\nonumber \\
&\times&
\frac{wJ_{2a}(z{\Sigma})J_{2a-1}( w{\Sigma})
-zJ_{2a-1}(z{\Sigma})J_{2a}(w{\Sigma})}{(zw)^{2a}(z^2-w^2)}.
\ee
The parameter $a$ is defined by
\be
a = N_f -\frac 12 +\frac {\nu}2.
\ee
Recently, the Bessel kernel got a great deal of attention in the literature.
For details and further references we refer to
\cite{FORRESTER-1993,TRACY-WIDOM-1993}.

\vskip 1.5cm
\noindent{\bf 5. Numerical results}
\vskip 0.5 cm
All results have been obtained for an ensemble of 32 instantons and
32 anti-instantons confined to a box of $2.37^3 \times 4.74
\Lambda_{QCD}^{-4}$.
For $N_f= 0$ averages were carried out for an ensemble of
100,000 configurations, whereas for $N_f =1,\, 2$ the average was over
20,000 configurations. In the latter case, configurations
 are correlated due
to the Monte-Carlo evolution, so that, effectively,
the size of the ensemble is somewhat smaller.

As stated in the introduction, our hypothesis is that the scale of the
microscopic fluctuations and the scale of the secular variations are
independent.
The two scales are separated via  a procedure called unfolding.
This procedure is not new. It has been widely used in the analysis of spectra
of quantum systems in terms of random matrix ensembles
\cite{BOHIGAS-GIANNONI-1984,SELIGMAN-VERBAARSCHOT-ZIRNBAUER-1984}.
It is always the
unfolded spectrum that has universal properties.
In the present case, we first obtain the average spectral density $\bar\rho$,
by fitting a smooth curve to the histogram of the eigenvalues
of the Dirac operator. In our case we use a combination of a gaussian
and an exponential function. This function is then
used to generate the unfolded spectrum  $\{\lambda^U_n\}$
from the original spectrum $\{\lambda_n\}$ according to
\be
\lambda^U_n = \int_0^{\lambda_n} \bar\rho(\lambda) d\lambda.
\ee
The average spacing of the unfolded spectrum is unity.

In Figs. 1 and 2 the solid lines show histograms of the unfolded spectrum
$\{\lambda^U_n\}$ for $N_c = 2$ and $N_c =3$, respectively. These results
are compared to
the microscopic spectral density for the chGUE (dashed curve)
and the chGOE (dotted curve). Results are given for 0, 1 and
2 flavors. In all cases, we find that the spectrum for $N_c = 2$ is described
by the chGOE, whereas the spectrum for $N_c = 3$ is described by the chGUE.
The only discrepancy is that  the
oscillations for $\beta=2$ are not reproduced several level spacings away from
zero. Since the positions of the peaks give the average location  of the
eigenvalues, this implies that level fluctuations are larger
than expected
according to random matrix theory. At this moment it is not clear
whether this effect is a finite size effect that will go away in the
thermodynamic limit.
For more than two flavors the gap near zero becomes so
big that the microscopic and the macroscopic scales can no longer be
separated unambiguously.

\vskip 1.5cm
\noindent{\bf 6. Discussion}
\vskip 0.5 cm
We have calculated the spectrum of the Dirac operator for an instanton liquid
approximation to the QCD partition function for both two and three colors.
For 0, 1 and 2 flavors, we have have found
that the spectrum near zero virtuality is different
in both cases. In all cases the microscopic limit
of the spectrum, obtained in the thermodynamic limit
by rescaling the eigenvalues $\sim V_4$,
can be described by chiral random matrix theory
with only the symmetries of the Dirac operator as input.
For two colors the Dirac operator is real and the appropriate ensemble, the
chiral orthogonal ensemble, has real matrix elements. For three colors
the Dirac operator is complex. Now the appropriate ensemble, the chiral
unitary ensemble, has complex matrix elements.
This confirms our universality hypothesis that
the spectrum near zero virtuality is entirely determined by symmetries.
In a sense, this is natural because this region of the spectrum carries
information on the very long wavelength excitations which are determined
by the global symmetries of the QCD partition function. Indeed, the
microscopic correlation functions are consistent with the static limit of
the QCD partition function.

As has been argued in \cite{VERBAARSCHOT-1994P}, there is a third
universality class, where the matrix elements of the chiral
random matrix ensemble are quaternion real. This case is realized for
fermions in the adjoint representation. It would of great interest
to subject this case to numerical studies as well.

\vglue 0.6cm
{\bf \noindent  Acknowledgements \hfil}
\vglue 0.4cm
The reported work was partially supported by the US DOE grant
DE-FG-88ER40388. We acknowledge the NERSC at Lawrence Livermore where
most of the computations presented in this paper were performed.

\vfill
\eject
\newpage
\setlength{\baselineskip}{15pt}

\bibliographystyle{aip}

\vfill
\eject
\setlength{\baselineskip}{21pt}
\noindent
{\bf Figure Captions}
\vskip 0.5cm
\noindent
Fig. 1. The microscopic spectral density $\rho_S(z)$ versus the virtuality
$z$ measured in units of the average level spacing.
The results are for the color gauge group SU(2).
The full line
shows the histogram of our numerical calculations. Analytical results
for the chGOE and the chGUE are represented by dotted and dashed
curves, respectively. For the number of flavors we refer to the labels
in the figure.
\vskip 1cm
Fig. 2. The microscopic spectral density $\rho_S(z)$ versus the virtuality
$z$ measured in units of the average level spacing.
The results are for the color gauge group SU(3).
Further explanation can be found in the caption of Fig. 1.


\end{document}